\documentclass[conference]{IEEEtran}

\usepackage{amsmath,graphicx,bm,bbm,algorithm,color} 
\usepackage{amsmath, mathtools, amsfonts, bm, amssymb, amsthm,graphicx, caption, subcaption,multirow}
\usepackage{algorithm} 
\usepackage[colorlinks=true, allcolors=black]{hyperref}

\usepackage{algpseudocode,matlab-prettifier}

\newcommand{\ik}{{ki}}

\newcommand{\x}{\bm{x}}

\renewcommand{\b}{\bm{b}}
\newcommand{\y}{\bm{y}}

\newcommand{\U}{{\bm{U}}}
\newcommand{\V}{{\bm{V}}}
\newcommand{\X}{\bm{X}}
\newcommand{\R}{{\bm{R}}}
\newcommand{\B}{\bm{B}}

\newcommand{\I}{\bm{I}}
\newcommand{\Y}{\bm{Y}}

\newcommand{\s}{\bm{s}}

\newcommand{\xhat}{\bm{{x}}}
\newcommand{\bhat}{\bm{{b}}}
\newcommand{\Uhat}{\U}
\newcommand{\Bhat}{\B}
\newcommand{\Xhat}{\X}

\newcommand{\Z}{{\bm{Z}}}

\newcommand{\z}{\bm{z}}

\newcommand{\A}{\bm{A}}

\newcommand{\M}{\bm{M}}

\renewcommand{\S}{\bm{S}}

\newcommand{\Ustar}{\U^*{}}

\newcommand{\Vstar}{\V^*{}}
\newcommand{\Xstar}{{\X^*}}
\newcommand{\xstar}{\x^*}

\newcommand{\bSigma}{{\bm\Sigma^*}}

\newcommand{\sigmin}{{\sigma_{\min}^*}}
\newcommand{\sigmax}{{\sigma_{\max}^*}}

\newcommand{\bstar}{\b^*}

\newcommand{\Bstar}{{\B^*}}

\newtheorem{theorem}{Theorem}[section]

\newtheorem{assu}[theorem]{Assumption}

\renewcommand{\subsubsection}[1]{\noindent {\em{#1. }}}

\newcommand{\vsm}{\vspace{-0.1in}}

\newcommand{\Sstar}{\S^*}

\newcommand{\bea}{\begin{eqnarray}} \newcommand{\eea}{\end{eqnarray}}
\newcommand{\ben}{\begin{enumerate}} \newcommand{\een}{\end{enumerate}}

\renewcommand{\L}{\bm{L}}
\renewcommand{\S}{\bm{S}}
\renewcommand{\l}{\bm{l}}
\renewcommand{\s}{\bm{s}}

\title{A Fast Algorithm for  Low Rank + Sparse \\column-wise Compressive Sensing}
\author{Silpa Babu,  Namrata Vaswani \\  
ECE dept, Iowa State University, USA }

\begin{document}
%
\maketitle

\renewcommand{\bhat}{\b}
\renewcommand{\xhat}{\x}
\renewcommand{\Xhat}{\X}
\begin{abstract}
This paper focuses studies the following low rank + sparse (LR+S) column-wise compressive sensing problem. We aim to recover an $n \times q$ matrix, $\X^* =[ \x_1^*, \x_2^*, \cdots , \x_q^*]$ from $m$ independent linear projections of each of its $q$ columns, given by $\y_k :=\A_k\x_k^*$, $k \in [q]$. Here, $\y_k$ is an $m$-length vector with $m < n$. We assume that the matrix $\X^*$ can be decomposed as $\X^*=\L^*+\S^*$, where $\L^*$ is a low rank matrix of rank $r << \min(n,q)$ and $\S^*$ is a sparse matrix. Each column of $\S$ contains $\rho$ non-zero entries. The matrices $\A_k$ are known and mutually independent for different $k$.
To address this recovery problem, we propose a novel fast GD-based solution called AltGDmin-LR+S, which is memory and communication efficient. We numerically evaluate its performance  by conducting a detailed simulation-based study. 
\end{abstract}



\section{Introduction}
The modeling of an unknown matrix as the sum of a low-rank (LR) matrix and a sparse matrix is widely employed in various dynamic imaging applications - foreground-background separation from videos \cite{rpca,rrpcp_review}, dynamic MRI reconstruction \cite{st_imaging,dyn_mri1}, and low-rank sketching \cite{lee2019neurips} are three important example applications. The design of recommendation systems with outliers is another one \cite{rpca,rrpcp_review}.
The low-rank component characterizes the static and slowly changing background across the frames, while the sparse component represents the sparse foreground -- moving objects in the case of videos, brain activation patterns in case of functional MRI, and motion of the contrast agents across the organs in case of contrast enhanced dynamic MRI.


\subsection{Problem and Notation}
We study the problem of recovering an $n \times q$ matrix $\X^*$ from undersampled measurements of each of its columns, i.e., from
\[
\y_k=\A_k \x_k^*, \hspace{0.5 cm} k \in [q]
\]
Here, $\A_k$ is a known matrix of dimension $m \times n$, with $m < n$. The different $\A_k$s are i.i.d. and each is a dense matrix: either a random Gaussian matrix (each entry i.i.d. standard Gaussian) or a random Fourier matrix (a random subset of rows of the discrete Fourier transform matrix).
The vectors $\y_k$ and $\xstar_k$ denote the $k$-th column of matrix $\Y$ and $\X^*$, respectively.

We assume that $\X^*=\L^*+\S^*$, where $\L^*$ is a low rank matrix and $\S^*$ is a sparse matrix. The rank of matrix $\L^*$ is denoted by $r$, and we assume $r \ll \min(n,q)$ (low-rank). Each column of $\S^*$ has $\rho$ non-zero entries. There is no bound on the magnitude of the non-zero entries of matrix $\S^*$. Therefore, the undersampled Gaussian measurements can be expressed as
\[
\y_k=\A_k (\l^*_k+\s^*_k), \ \   k \in [q]
\]
where $\l^*_k$ and $\s^*_k$ is the k-th column of  matrix $\L^*$ and $\S^*$ respectively.
The above problem occurs in dynamic MRI ($\A_k$s are a subset of rows of the 2D-DFT matrix) or in sketching ($\A_k$'s are random Gaussian) \cite{lee2019neurips}

The reduced (rank $r$) Singular Value Decomposition of matrix $\L^*$ can be represented as $\U^* \bSigma \V^*$. The matrix $\L^*$ can be written as
\[
\L^*=\U^* \B^*
\]
where $\U^*$ is an $n \times r$ matrix with orthonormal columns, and $\B^*$ is an $r \times q$ matrix given by $\B^*=\bSigma \V^*$. Consequently, each column of $\X^*$ is given by $\x^*_k=\U^*\b^*_k+\s^*_k$.
Hence the undersampled Gaussian measurements can be written as
\[
\y_k=\A_k (\U^*\b^*_k+\s^*_k), \hspace{0.5 cm}  k \in [q]
\]
We use $\|. \|$ without a subscript denotes the $l_2$ norm of a vector or the induced $l_2$ norm of a matrix, and we use $\|.\|_F$ to denote the matrix Frobenius norm. For a tall matrix $\M$, $\M^\dag:=(\M^{\top}\M)^{-1}\M^{\top}$ denotes its Moore-Penrose pseudo-inverse.

We studied the $\S^*=0$ special case of the above problem in our recent work \cite{lrpr_gdmin,lrpr_gdmin_mri}.
Recall that
	$
	\L^* \stackrel{SVD}{=} \Ustar  \bSigma \Vstar: = \Ustar \Bstar
	$
	denote its reduced (rank $r$) SVD, and $\kappa:= \sigmax/\sigmin$ the condition number of $\bSigma$.
	Notice that our problem is asymmetric across rows and columns. Also, each measurement $\y_\ik$ is a global functions of column $\x^*_k$, but not of the entire matrix. Hence, we need the following assumption, which is a subset of the assumptions used in the LR matrix completion literature. 

\begin{assu}
\label{right_incoh}
We assume that $\max_k \| \bstar_k \| \leq \mu \sqrt{r/q} {\sigmax}^2$ for a constant $\mu \geq 1$. 
\end{assu}

\subsection{Related Work}
To our best knowledge, the above problem has only been studied in the dynamic MRI setting \cite{otazo2015low, lin2018efficient, 6808502, 7831389}. 
All these algorithms are extremely slow and  memory-inefficient because they requires storing and processing the entire matrix $\L$ (these do not factorize $\L$ as $\L = \U \B$). Moreover, all these works only focused on accurately recovering real MRI sequences from the available measurements. To our knowledge, none of these do a careful simulation-based or theoretical study of when the proposed algorithm converges and why.
Another related work from the MRI literature is \cite{lingala2011accelerated}; this explored the LR\&S model for MRI datasets.

While the above problem has not been explored in the theoretical or signal processing literature, related LR+S problems have been extensively studied. The robust PCA problem is the above problem with $\A_k = \I$, i.e. the goal is to separate $\L^*$ and $\S^*$ from $\X^*:=\L^*+\S^*$. The robust LR matrix completion (LRMC) problem, also sometimes referred to as partially observed robust PCA, involves recovering $\L^*$ from a subset of the entries of $\Xstar$. This can be understood as the above problem with $\A_k$ being a 1-0 matrix with exactly one 1 in each row. Provably correct convex \cite{rpca,rpca2}, alternating minimization \cite{robpca_nonconvex}, and GD-based solutions to robust PCA and robust LRMC \cite{rpca_gd,rmc_gd} have been developed in the literature. The work of \cite{rpca_gd} factors $\L$ as $\L = \U \B$ and uses an alternating GD (AltGD) algorithm that alternatively updates $\U, \B,\S$. Since $\L = \U \B = \U \R \R^{-1} \B$  for any $r \times r$ invertible matrix $\R$ (the decomposition is not unique), the alternating GD algorithm requires adding a term to the cost function that ensures the norms of $\U$ and $\B$ are balanced. Thus, this work develops an AltGD algorithm that minimizes $\sum_k \|\y_k - \A_k \U \b_k - \A_k \s_k\|^2 + \|\U^\top \U - \B \B^\top\|_F$.

Another related problem is that of recovering $\X^*:=\L^*+\S^*$ from dense linear projections of the entire matrix $\Xstar$, i.e., from $\y_j:= \mathcal{A}_j(\Xstar)$, $j=1,2,\dots,j_{max}$ (this can be referred as low-rank plus sparse compressive sensing) \cite{tanner2023compressed}. Here $\mathcal{A}_j(\Xstar):= \langle \A_j, \Xstar \rangle$ with $\A_j$ being random Gaussian. Notice that this problem is different from ours where we have dense linear projections of each column of $\Xstar$ but not of the entire matrix.

In our recent work \cite{lrpr_gdmin,lrpr_gdmin_mri,lrpr_gdmin_mri_jp}, we studied the LR column-wise compressive sensing (LRCCS) problem which is the above problem with $\Sstar = \bm{0}$. This is also where we introduced the alternating GD and minimization (AltGDmin) algorithm idea. In \cite{lrpr_gdmin}, we provided theoretical sample complexity and iteration complexity guarantees for it and extensive numerical simulations. In \cite{lrpr_gdmin_mri,lrpr_gdmin_mri_jp}, we developed a practical modification for fast accelerated dynamic MRI reconstruction. Via extensive experiments on real MRI image sequences (with simulated Cartesian and radial sampling) and on real scanner data, we showed that AltGDmin-MRI is both significantly faster and more accurate than many state-of-the-art approaches from MRI literature.

\subsection{Contributions}
We develop a novel GD-based algorithm called Alternating GD and minimization or AltGDmin for solving the above LR+S column-wise compressive sensing problem (LR+S-CCS) and argue why it is both fast and memory and communication-efficient. The communication-efficiency claim assumes a federated setting in which subsets of $\y_k,\A_k$ are available at  different distributed nodes. Using extensive simulation experiments, we also numerically evaluate the sample complexity (value of $m$ needed for a given $n$, $q$, $r$, $\rho$) to guarantee algorithm convergence.

The AltGDmin algorithmic framework was first introduced for solving the LRCCS problem in \cite{lrpr_gdmin}. Its idea is to split the unknown variables into two parts, $\Z_a, \Z_b$ and alternatively update each of them using GD or projected GD for $\Z_a$ and minimization for $\Z_b$. We let $\Z_b$ be the subset of variables for which the minimization can be ``decoupled'', i.e., subsets of $\Z_b$ are functions of only a subset of $\Y$.
In problems such as LRCCS or the above LR+S-CCS problem, the decoupling is column-wise. Assume a factored representation of $\L$, i.e.,  $\L = \U \B$ where $\U$ and $\B$ are matrices with $r$ columns and rows respectively.
In our current problem, $\Z_b \equiv \{\B,\S \}$ and $\Z_a = \U$.  Thus, our algorithm alternates between a projected GD step for updating $\U$ and a minimization step for updating $\B$ and $\S$.

Because of the column-wise decoupling, the minimization step is about as fast as the GD step. As we argue later, the per-iteration time complexity is $mqnr$, memory complexity is $\max(n,q)\cdot \max(r,\rho)$ and the communication complexity is $nr$ per node. 
Memory and communication cost wise our algorithm is much better than projected GD. Time-wise it is faster than AltMin which solves a minimization problem also for updating $\U$. This problem is coupled across all columns and hence is a much more expensive problem to fully re-solve at each algorithm iteration. Replacing this by a single GD update is what makes AltGDmin much faster than AltMin.

To our knowledge, this work is the first to do a detailed simulation-based study of the above LR+S column-wise compressive sensing problem. As is well known from older compressive sensing literature for sparse recovery problems, as well as for LRCCS, algorithms developed originally for the random Gaussian $\A_k$ setting also often work with simple modifications for solving the MRI reconstruction problem, e.g., \cite{candes,donoho} and \cite{sparsemri} for CS, and \cite{lrpr_gdmin} and \cite{babu2023fast} for LRCCS.

\section{AltGDmin for L+S column-wise compressive sensing (L+S-CCS)}

We are interested in developing a fast gradient descent (GD) based algorithm to find matrices $\L$ and $\S$ that minimize the cost function:
\[
f(\L ,\S) =\sum_{k=1}^q \|\y_k-\A_k (\l_k+\s_k)\|^2
\]
subject to the constraints (i) matrix $\L$ has rank $r$ or less, (ii) each column of matrix  $\S$ is $\rho$ sparse. This means that the number of non-zero entries in each column $\s_k$ is at most $\rho$. As previously mentioned, we can impose the rank constraint implicitly by expressing the matrix $\L$ as $\L=\U \B$ where $\U$ and $\B$ are $n \times r$ and $r \times q$ matrices. The cost function can then be written as:
\[
f(\U, \B ,\S) =\sum_{k=1}^q \|\y_k-\A_k (\U\b_k+\s_k)\|^2
\]
Our objective is to find matrices $\U$, $\B$, and $\S$ that minimize the cost function $f(\U, \B ,\S)$ while satisfying the constraint that each column of $\S$ is $\rho$ sparse.
\subsection{altGDmin-L+S iteration}
The gradient of the cost function with respect to $\U$  is given by $ \nabla_U f (\Uhat, \Bhat, \S) $.
\[
\nabla_U f (\Uhat, \Bhat, \S) = \sum_{k=1}^q \A_k^\top (\A_k( \Uhat \bhat_k+  \s_k)- \y_k) \bhat_k^\top.
\]

At each new iteration, the following steps are performed:
\begin{itemize}
\item The matrix $\U$ is updated using projected gradient descent.
\item For a given $\U$, the matrices $\S$ and $\B$ are updated by using minimization, as explained in detail in Section \ref{B}.

\end{itemize}
These steps iteratively update ${\U}$, ${\B}$, and ${\S}$ to minimize the cost function and obtain better estimates. We summarize our proposed algorithm altGDmin-L+S in Algorithm \ref{lpluss_IHTmin}.

\subsection{Minimization: Updating S and B for a given U}
\label{B}
For a given $\U$, we need to find $\s_k$ and $\b_k$ that minimize $f(\U, \B,\S) =\sum_{k=1}^q \|\y_k-\A_k (\U\b_k+\s_k)\|^2$. Observe that $\s_k,\b_k$ appear only in the $k$-th term. Thus,
\[
\min_{\B,\S} f(\U, \B,\S) = \sum_{k=1}^q \min_{\s_k, \b_k} \|\y_k-\A_k (\U\b_k+\s_k)\|^2
\]
This means that the minimization problem can be decoupled across each columns, making it much faster.
Moreover, notice that, for a given $s_k$, the recovery of $b_k$  is a standard least squares (LS) problem with a closed-form solution that can be written in terms of $\s_k$:
\begin{align}
\b_k = (\A_k  \U)^\dagger (\y_k- \A_k \s_k)
\label{bk_LS}
\end{align}
Recall here that $\M^\dagger=(\M^{\top}\M)^{-1}\M^{\top}$. We can substitute this for $\b_k$ and hence get a minimization over only the $\s_k$s, i.e.,
\[
 \sum_{k=1}^q \min_{\s_k, \b_k} \|\y_k-\A_k (\U\b_k+\s_k)\|^2 =  \sum_{k=1}^q \min_{\s_k} \|\z_k- \M_k\s_k\|^2
\]
%
where $\z_k:=\y_k-\A_k\U(\A_k  \U)^\dagger \y_k$ and $\M_k:=\A_k-\A_k\U(\A_k  \U)^\dagger \A_k$. In summary,
\[
\min_{\B,\S} f(\U, \B,\S) = \sum_{k=1}^q \min_{\s_k} \mathcal{L}(\s_k), \ \mathcal{L}(\s_k):= \|\z_k- \M_k\s_k\|^2
\]
with $\z_k, \M_k$ as given above. Thus, we have simplified the min over $\B,\S$ into a problem of solving $q$ individual problems, each of which is a standard sparse recovery problem, $\min_{\s_k} \mathcal{L}(\s_k)$.
Various algorithms can be used to solve the sparse recovery problem. In this paper, we choose to use the Iterative Hard Thresholding (IHT) algorithm provided in \cite{blumensath2009} due to simplicity and efficiency.
For a given $\U$ and $\S$, the matrix $\B$ can be updated  using \eqref{bk_LS}.

\begin{algorithm}[t]
\caption{\sl{altGDmin-L+S: Low Rank plus Sparse Model. Let $\M^\dagger:= (\M^\top\M)^{-1} \M^\top$.}}
\label{lpluss_IHTmin}
\begin{algorithmic}[1]
   \State {\bfseries Input:} $\y_k, \A_k, k \in [q]$.

\State {\bf Initialization:} {Set  $\tau_{0max}=10$.}
{\small

 \State $\s_k$=IHT($\y_k, \A_k, \rho, \tau_{0max}, \bm{0}$), $k \in [q]$

\State
\begin{align*}
\L_0 & :=   [   \A_1^\top (\y_{1}-\A_1 \s_1),  ...,  \A_k^\top (\y_{k}-\A_k \s_k), ...,\A_q^\top (\y_{q}-\A_q \s_q)]
\end{align*}

\State  Set $\Uhat_0 \leftarrow $ top $r$ left singular vectors of $\L_{0}$.
\State  Let $\Uhat \leftarrow \Uhat_{0}$.
 \State$\bhat_k \leftarrow  (\A_k  \U)^\dagger (\y_k- \A_k \s_k)$ , $k \in q$.
}
\State {\bf GDmin iterations:} { Set $T_{max} = 120$, $\tau_{max}=3$, set $\eta=\frac{0.14}{\|\nabla_U f (\Uhat_0, \Bhat_1, \S_1) \|}$} 
  \For{$t=1$ {\bfseries to} $T_{max}$}
  \State Compute:
   $ (\z_k)_t=\y_k-\A_k\U(\A_k  \U)^\dagger \y_k$ and \\
   $(\M_k)_t=\A_k-\A_k\U(\A_k  \U)^\dagger \A_k$,  $k \in [q]$
	 \State $(\s_k)_t$=IHT($(\z_k)_t, (\M_k)_t, \rho, \tau_{max}, (\s_k)_{t-1}$), $k \in [q]$
	 \State$(\bhat_k)_t   \leftarrow  (\A_k  \U)^\dagger (\y_k- \A_k (\s_k)_t)$ , $k \in q$.
	  \State Gradient compute:
  $$\nabla_U f (\Uhat, \Bhat, \S) \leftarrow \sum_{k=1}^q \A_k^\top (\A_k( \Uhat (\bhat_k)_t+  (\s_k))_t)- \y_k) (\bhat_k)_t^\top$$

  \State   Projected GD step :
 \small{ $\Uhat^+   \leftarrow QR(\Uhat - \eta \nabla_U f(\Uhat, \Bhat, \S))$}.
  \State Set $\Uhat\leftarrow \U^+$.

\EndFor
\State      {\bfseries Output:} $\Xhat:= [\xhat_1, \xhat_2, \dots, \xhat_q],$ where $\x_k=\U (\b_k)_T + (\s_k)_T$.

\end{algorithmic}
\end{algorithm}

\subsection{altGDmin-L+S initialization}
Since the cost function $f(\U,\B,\S)$ is non-convex, it needs a careful initialization. As in all previous work on iterative algorithms for the most related L+S problem, robust PCA, we initialize $\S$ first while letting $\B= \bm{0}$.
The reason this is done is there is no upper bound on entries of the sparse matrix which usually models large magnitude, but infrequently occurring, outlier entries. On the other hand, when we assume that the condition number of $\L$ is a numerical constant, we are implicitly assuming an upper bound on the entries of $\L$.
Also, as noted earlier, the recovery of $\s_k$'s is decoupled across columns. \\
Given an initialize estimate of $\S$, we initialize $\U$ by a standard spectral initialization approach: compute $\U$ as the top $r$ singular vectors of the matrix
\[
\L_0:= [   \A_1^\top (\y_{1}-\A_1 \s_1),  ...,  \A_k^\top (\y_{k}-\A_k \s_k), ...,\A_q^\top (\y_{q}-\A_q \s_q)]
\]

\subsection{Iterative Hard Thresholding}

 At each iteration of IHT \cite{blumensath2009}, a vector $\s$ is updated as follows:

\begin{itemize}
\item Compute gradient of the cost function with respect to $\s$.
\item Compute the step size using the gradient.
\item Update $\s$ using gradient descent.
\item Perform Hard Thresholding on the updated $\s$ to keep only the largest $\rho$ entries of $\s$ in magnitude.
\end{itemize}
We summarize IHT algorithm in Algorithm \ref{IHT}. The computational time for updating a vector $\s$ using the IHT algorithm in initialization mainly depends on the operator $\M$ and its transpose $\M^\top$. If these operators are general matrices, the computational time requirement is $O(mn)$ per iteration for each column. 

\begin{algorithm}[h]
\caption{\sl{IHT }}
\label{IHT}
\begin{algorithmic}[1]
   \State {\bfseries Input:} $\z, \M, \rho, \tau_{max}, \s$.

  \For{$t=1$ {\bfseries to} $\tau_{max}$}
  \State  Gradient Compute $\nabla_{\s} \mathcal{L}(\s)$, which is the gradient of $\mathcal{L}(\s)=\|\M \s-\z  \|^2 $ with respect to $\s$.
	 \State Compute step size $\mu=\frac{\|\nabla_{\s} \mathcal{L}(\s)\|}{\|\M\nabla_{\s} \mathcal{L}(\s)\|}$
	 \State GD step for $(\s)_t\leftarrow \s-\mu\nabla_{\s} \mathcal{L}(\s)$
	  \State $\s \leftarrow \mathrm{HardThreshold}$ $((\s)_t, \rho)$, which keep only the largest $\rho$ entries of $\s$ in magnitude.

\EndFor
\State      {\bfseries Output:} $\s$.

\end{algorithmic}
\end{algorithm}

\subsection{Time, Communication, and Memory Complexity}


\subsubsection{Time Complexity}
The time complexity of the IHT algorithm for recovering a vector $\s$ in the initialization is $O(mn)$ per iteration, as previously explained. Consequently, the time complexity of the IHT algorithm for updating the entire matrix $\S$  is $O(\tau_{0max}mnq)$. The initialization step requires a time complexity of $mqn$ for computing $\L_0$.
The time needed for computing the $r$-SVD of $\L_0$ is given by $nqr \times $(number of iterations of power method).
Summing up the individual time complexities mentioned above, the time complexity of the initialization step is given by $O(\tau_{0max}mqn+nqr\times $(number of iterations of power method)$+mqn)$. However, since the dominant term in this expression is $O(\tau_{0max}mqn)$, the overall time complexity of initialization is  $O(\tau_{0max}mqn)$.

The computational time required for updating each column of $\S$ using IHT is $O(mn)$ per iteration (explained earlier). The time complexity for $\S$ update using IHT algorithm is $O(\tau_{max}mqn)$, where $\tau_{max}$ is the maximum number of iterations in the IHT algorithm.
The update of columns of $\B$ by LS also needs time $O(mnqr)$.
One gradient computation with respect to $\U$ needs time $O(mqnr)$. The QR decomposition requires time $nr^2$.
We need to repeat these steps $T_{max}$ times. Thus, the total time complexity of altGDmin-L+S iterations  is $O((mnqr+nr^2+mnqr+\tau_{max}mqn)T_{max})=O(mnqrT_{max})$.

Therefore, the total time complexity of the altGDmin-L+S algorithm is $O(mnqrT_{max})$.

\subsubsection{Communication Complexity}
The communication complexity per node per iteration in a distributed implementation (where subsets of $\y_k$s are processed at different distributed nodes) is given as $O(nr)$. 

\subsubsection{Memory Complexity}
If each column of matrix $\S$ has only $\rho$ non-zero entries, then storing each column requires $2\rho$ numbers. Since $\S$ has $q$ columns, the total memory required to store $\S$ is $2\rho q$.
The memory required to store $n \times r$ matrix $\U$ and $r \times q$ matrix $\B$ is  $nr$ and $qr$. In comparison, storing the entire matrix $\X$ would require memory of $nq$. Therefore, by storing $\S$, $\U$, and $\B$ instead of the full matrix $\X$, the algorithm can significantly reduce the memory requirements. The memory complexity of storing $\U$, $\B$, and $\S$ is given by $\max(nr, q r,  q \rho)$.


\subsection{Settings parameters automatically}

In real-world applications, the rank $r$ of matrix ${\L^*}$ and the exact sparsity of each column in matrix $\S$ are unknown. To estimate a suitable value for $r$, the "$b\%$ energy threshold" method can be employed on the singular values of the initial estimate of matrix $\L$ denoted as $\L_0$. Additionally, the estimated value of $r$ should be sufficiently less than the minimum of $n$ and $q$ to ensure a low-rank property. To determine $r$, we perform the "$b\%$ energy threshold" on the first ${\min}\left(\frac{n}{10}, \frac{q}{10}, \frac{m}{10}\right)$ singular values. In our experiments, a $65\%$ energy threshold was used.
In practical scenarios, the exact value of $\rho$ (the sparsity level per column) is often unknown. In many cases the maximum possible nonzero entries (upper bound on $\rho$), denoted $\rho_{\max}$ is known.  We assume this in our code and experiments.
The maximum number of iterations of the IHT algorithm in the initialization  $\tau_{0max} $ is set to $10$.

For the altGDmin-L+S Algorithm, the maximum number of iterations $T_{max}$ is set to $200$. The step size $\eta$ for updating the matrix $\U$ is set to $\frac{0.14}{\|\nabla_U f (\Uhat_0, \Bhat_1, \S_1) \|}$, where $\U_0$ is the initial estimate, $\B_1$ and $\S_1$ are the estimates in the first AltGDmin-L+S iteration. Assuming that the gradient norm decreases over iterations, this implies that at any iteration $t$,  we have $\eta \nabla_U f (\Uhat_t, \Bhat_t, \S_t) < 1$. The maximum number of iterations for the IHT algorithm, denoted as $\tau_{max}$, used for updating $\S$ is set to $3$.

\section{Experiments}
The matrix $\U^*$ is generated by orthonormalizing an $n \times r$ matrix with independent and identically distributed (i.i.d) Gaussian entries. The matrix $\B^*$ and $\A_k$ are matrices of normally distributed random numbers of dimensions $r \times q$ and $m \times n$, respectively. The locations of $\rho$ non zero entries of each column of matrix $\S$, is sampled uniformly at random, without replacement, from the integers 1 to n. Each column of $\S^*$ contains exactly $\rho$ non-zero entries. 
In each of the 100 Monte Carlo runs, the measurement matrices $\A_k$ consist of i.i.d standard Gaussian entries. We obtained the Gaussian measurements as $\y_k=\A_k (\U^* \b^*_k+\s^*_k)$, $k \in [q]$. The error is computed as $Error(\X^*,\X)=\frac{\|\X^*-\X \|_F}{\|\X^*\|_F}$,where $\|\cdot\|_F$ denotes the Frobenius norm. For IHT implementation, we used the modified version of code provided in \cite{jabari2023}.

\begin{figure}[h]
\includegraphics[width=9cm,height=7 cm]{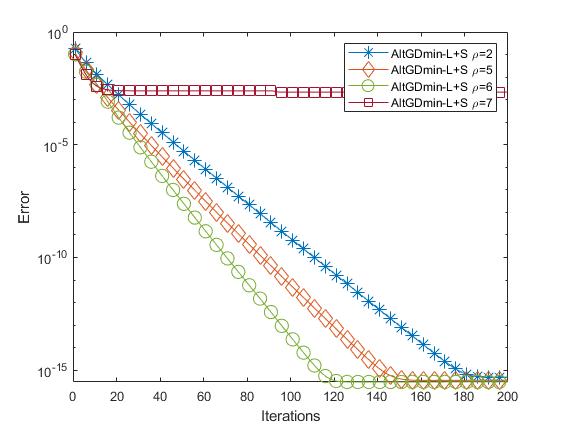}
\caption{\sl\small{{\bf Random Gaussian Measurements:}  We compare altGDmin-L+S performance for different values of $\rho$. Parameters used: $n=600$, $q=600$, $m=80$, $r=4$, $\rho_{\max}=7$, $T_{max} =200$.
}}
\label{FigAl1}
\end{figure}

For Fig. \ref{FigAl1},  the parameters  were set as follows: $T_{max}=200$,  $\tau_{0max}=10$, $\tau_{max}=3$, and $\eta=\frac{0.14}{\|\nabla_U f (\Uhat_0, \Bhat_1, \S_1) \|}$. In this experiment, we fixed the parameter values to $n=600$, $q=600$, $m=80$, $r=4$. Data was generated for different $\rho$ values, specifically $\rho=2, 5, 6$ and $7$. We considered $\rho_{\max}=7$. The values of each nonzero element of $\S^*$  were drawn uniformly from the interval $[-\alpha, \alpha]$, where $\alpha $ was chosen to be 6. This choice of values for the entries of $\S^*$ ensures that certain sparse components are smaller or larger than certain entries of $\L$, while some other components fall within the range of $\L$. This generation method for matrix $\S$ is referred to as S1 in this paper to simplify the explanation. On the y-axis, we display the empirical average of the error of matrix $\X$ using a semilog scale, which is averaged over 100 Monte Carlo simulations.  Notably, from Fig. \ref{FigAl1}, it can be observed that error converges to $10^{-15}$ in most cases. However, when $\rho$ is sufficiently large, our algorithm only converges to $10^{-3}$.

\begin{figure}[h]
\includegraphics[width=9cm,height=7 cm]{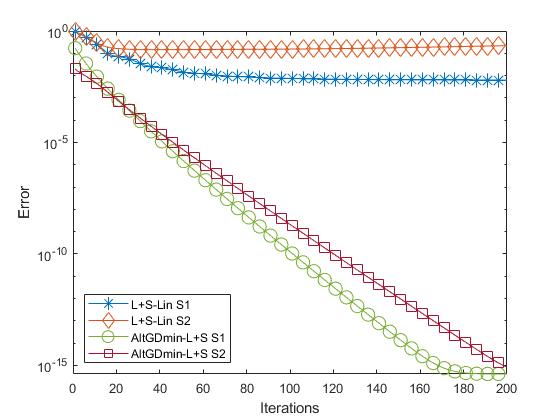}
\caption{\sl\small{{\bf Random Gaussian Measurements:} We compare the performance of altGDmin-L+S and L+S-Lin \cite{lin2018efficient} algorithm, using two different methods (S1 and S2) for generating nonzero entries Sparse matrix. Parameters used: $n=600$, $q=600$, $m=80$, $r=4$, $\rho=2$, $\rho_{\max}=5$, $T_{max} =200$.
}}
\label{FigAl2}
\end{figure}

In Fig. \ref{FigAl2}, we used the same set of parameters as in Fig. \ref{FigAl1}. The data was generated using the following parameter values: $n=600$, $q=600$, $m=80$, $r=4$, $\rho={2}$. We considered $\rho_{\max}=5$. In this experiment, the non-zero entries  of matrix $\S^*$ were generated using two methods: (1) S1, as explained earlier, and
(2) we referred second method as S2, where each nonzero element of  $\S^*$  was randomly chosen from the set $\{-1, -10, -100, 1,10,100\}$. 
In this experiment, our aim was to compare altGDmin-L+S with the well-known L+S-Lin \cite{lin2018efficient} algorithm in the MRI literature, using two different methods (S1 and S2) for generating nonzero entries matrix $\S$. To conduct this comparison, we used the code provided  by the authors for L+S-Lin \cite{lin2018efficient}, which is available on GitHub at \url{https://github.com/JeffFessler/reproduce-l-s-dynamic-mri/tree/main}; we used its version developed for dynamic MRI of abdomen. This code fails completely for our current simulated random Gaussian measurements' setting. In general too this algorothm needs application specific parameter tuning. To make it work, we made necessary modifications to the code to accommodate Gaussian measurements instead of Fourier measurements; and we provided it with our estimate of the rank of $\L$ (computed as explained earlier) and the value of $\rho_{\max}$. We compare its performance with our proposed algorithm, AltGDmin-L+S, in Fig. \ref{FigAl2}. 
With these changes, the recovery error of L+S-Lin goes down to about 0.0001 but the algorithm does not converge.  AltGDmin-L+S does converge. 

Next we compared our proposed algorithm with both L+S-Lin (its modified version explained above) \cite{lin2018efficient} and with our old algorithm altGDmin-LR \cite{lrpr_gdmin} designed for LRCCS problem for various values of $m$. We show the results in Table \ref{table-algos1}. The data was generated using the following parameter values: $n=400$, $q=400$, $r=4$, and  $\rho={2}$. We considered $\rho_{\max}=5$ and used the same parameter settings as in Fig. \ref{FigAl2}, except for $T_{max}$, which was changed to $10$. The matrix $\S$ was generated using method S1, and Fourier measurements were used, indicating that the matrices $\A_k$ were random Fourier measurements. We compare the performance of the altGDmin-L+S (proposed) algorithm with other two algorithms, L+S-Lin \cite{lin2018efficient} and altGDmin-LR \cite{lrpr_gdmin}, for different values of $m$. From table, it is clear that the altGDmin-L+S algorithm performs better than L+S-Lin and altGDmin-LR in Fourier settings. 

\begin{table}
\begin{center}
{
\begin{tabular}{|c|c|c|c|c|}
\hline
m & AltGDmin & L+S-Lin & AltGDmin-L+S   \\
\hline
40 &   1&   0.6986 &   0.4299  \\
60 &   1 &   0.5054 &   0.0366  \\
80 &   1&   0.3153 &   0.0093  \\
100 &   1 &   0.1749 &   0.0037  \\
150 &   1&   0.0685 &   0.0005  \\
200 &   0.9957 &   0.0739 &   0.0001  \\
250 &   0.9824 &   0.0509 &   0.0000  \\
300 &   0.9691 &   0.0281 &   0.0000  \\
\hline
\end{tabular}
}
\vsm
\end{center}
\caption{\sl\small{ {\bf Random Fourier Measurements:}
We report {\bf Error} for algorithms AltGDmin \cite{lrpr_gdmin}, L+S-Lin \cite{lin2018efficient} and AltGDmin-L+S algorithms for different values of m. 
Parameters used: $n=400$, $q=400$, $r=4$, $\rho=2$, $\rho_{\max}=5$, $T_{max} =10$.}}
\vsm
\label{table-algos1}
\end{table}

\section{Conclusions}
We developed a fast, memory and communication-efficient gradient descent (GD) based recovery algorithm,
called altGDmin-L+S for the LR+S-CCS  problem. Through simulations, we have shown that altGDmin-L+S achieves successful recovery of simulated data from their sketches. In future work, we plan to evaluate the performance of altGDmin-L+S in real-time video applications.  

\bibliographystyle{IEEEtran}
\bibliography{../../bib/tipnewpfmt_kfcsfullpap,./refs_Silpa}

\end{document}